\documentclass{webofc}
\usepackage[varg]{txfonts}   % Web of Conferences font
\usepackage{xcolor}
\newcommand{\ee}{e^{+}e^{-}}

\newcommand{\jp}{J/\psi}

\newcommand{\pipi}{\pi^{+}\pi^{-}}

\newcommand{\piz}{\pi^{0}}
\newcommand{\bbbar}{b\bar{b}}

\newcommand{\ccbar}{c\bar{c}}

\newcommand{\rt}{\rightarrow}

\newcommand{\jpsi}{J/\psi}

\newcommand{\dstrz}{D^{*0}}

\newcommand{\dstrbar}{\bar{D}^{*}}
\newcommand{\dstrzbar}{\bar{D}^{*0}}
\newcommand{\dz}{D^{0}}

\newcommand{\dzbar}{\bar{D^{0}}}

\newcommand{\chiczerop}{\chi_{c0}^{\prime}}

\newcommand{\chictwop}{\chi_{c2}^{\prime}}

\begin{document}
\title{The XYZ mesons: what they aren't}
%
% subtitle is optional
%
%%%\subtitle{Do you have a subtitle?\\ If so, write it here}

\author{\firstname{Stephen Lars} \lastname{Olsen}\inst{1}\fnsep\thanks{\email{solsen@ucas.ac.cn}}
}

\institute{Department of Physics, University of the Chinese Academy of Science, Beijing 100049, CHINA}

\abstract{I discuss the properties of some representative $XYZ$ mesons in the context of the most
commonly proposed models for their underlying nature.
}
\maketitle
\section{Some recent history}
\label{history}
\begin{figure}[htb]
\begin{minipage}[t]{70mm}  \includegraphics[height=0.75\textwidth,width=0.75\textwidth]{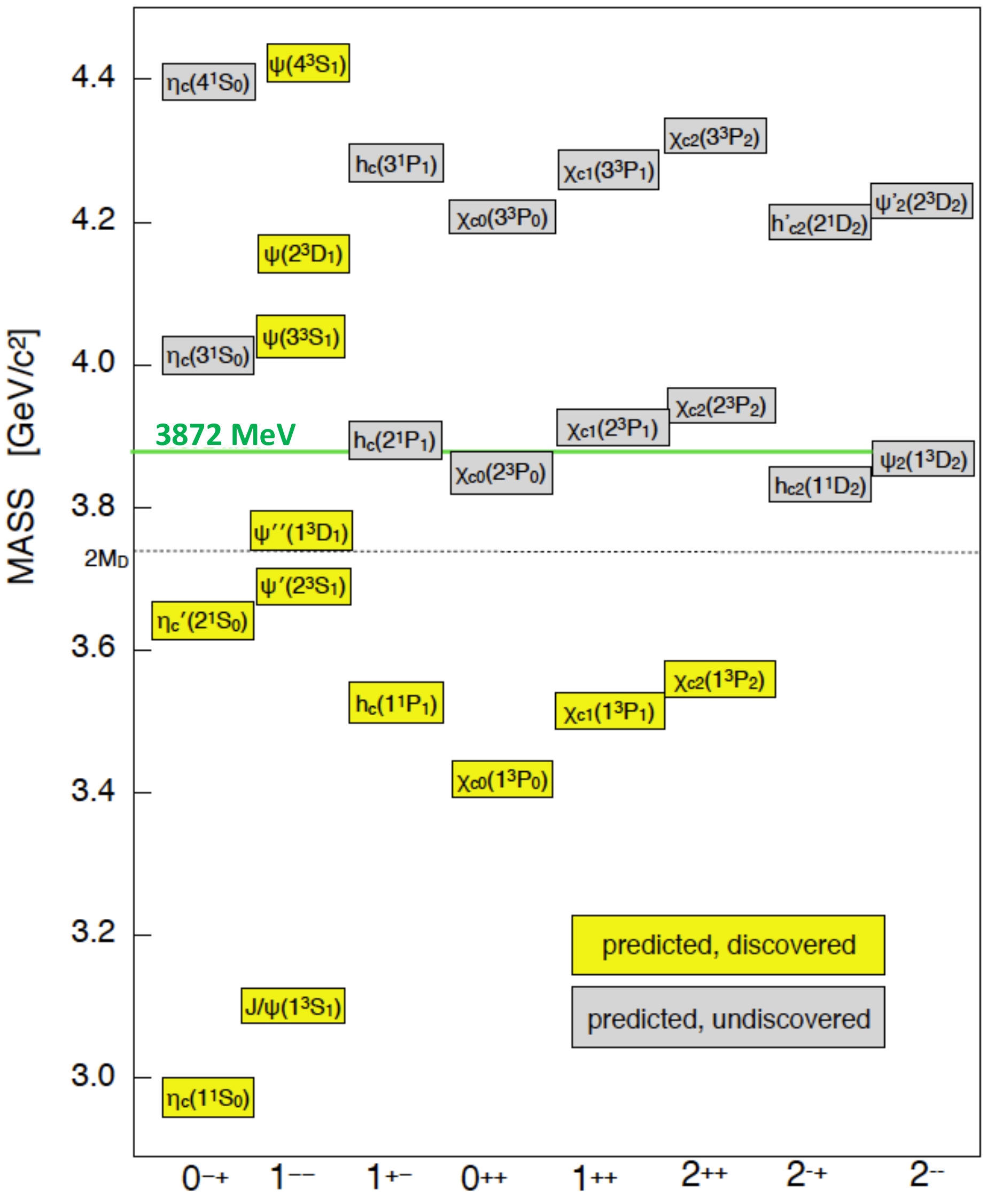}
\caption{\footnotesize The charmonium spectrum circa 2003.
}
\label{fig:charmonium}
\end{minipage}
\begin{minipage}[t]{17mm}
\end{minipage}
\begin{minipage}[t]{70mm}
  \includegraphics[height=0.75\textwidth,width=0.75\textwidth]{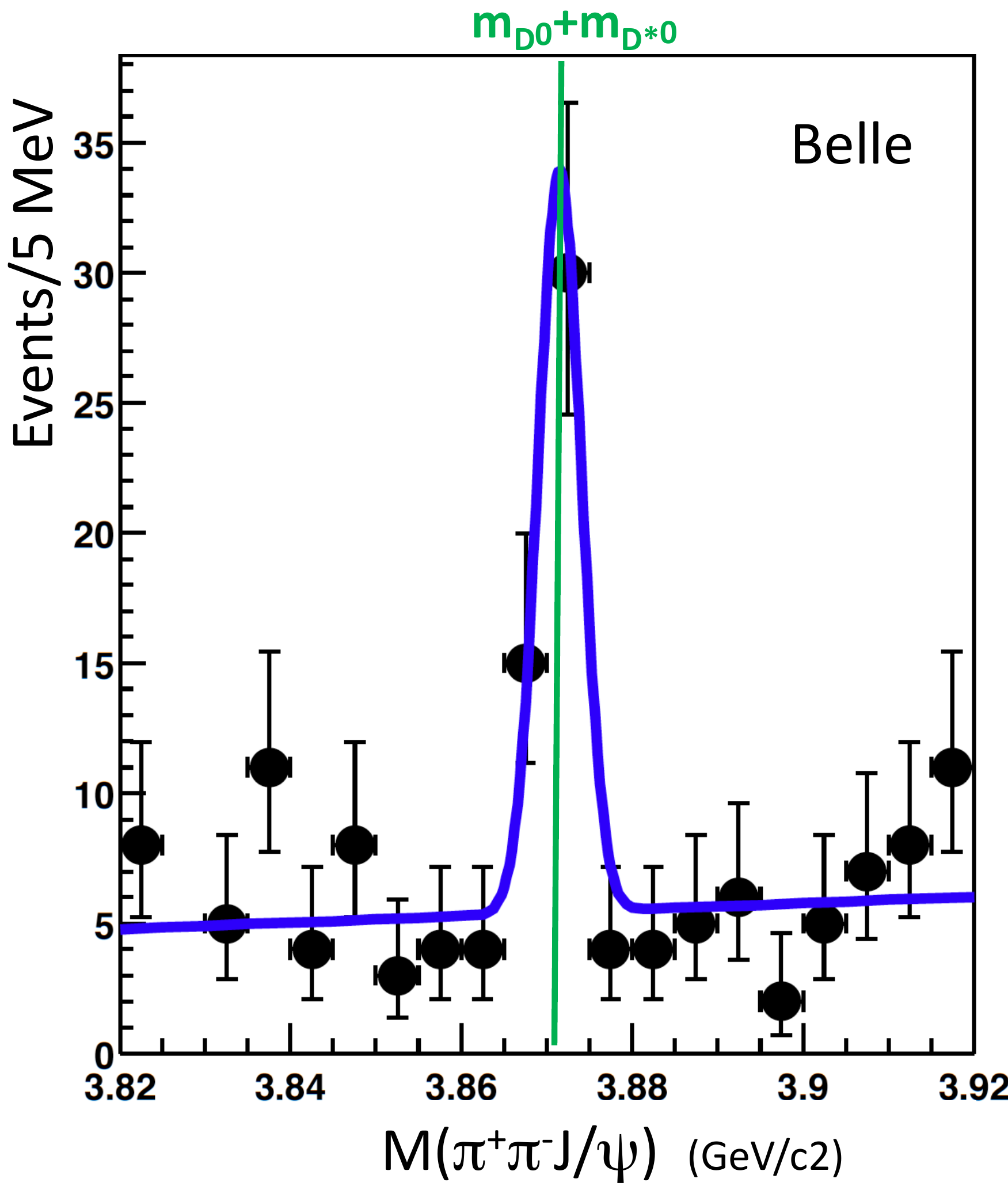}
%\end{minipage}\hspace{\fill}
  \caption{\footnotesize The $X(3872)$ signal with all the Belle data~\cite{Choi:2011fc}.
  }
\label{fig:x3872}
\end{minipage}\hspace{\fill}
\end{figure}

  The study of hadron spectroscopy had enormous success in the latter part of the
  twentieth century, when the charmonium ($\ccbar)$ and bottomonium ($\bbbar$)
  mesons were discovered and it was established that the mass spectra of these states, and
  many of their properties, could be accurately described by the Quarkonium model, which is
  based on non-relativistic Quantum Mechanics with a simple potential comprised
  of a coulombic short-ranged component smoothly coupled to a linearly rising
  ``confining'' term at larger distances.  Figure~\ref{fig:charmonium} shows
  the status of the charmonium spectrum in 2003, where the established charmonium
  mesons are colored yellow and the predicted but at that time unassigned states
  are gray. The assigned mesons all have properties that closely match their
  model-based expectations. Moreover, exceptions, {\it i.e.}, $\ccbar$
  mesons that could not be accommodated by this simple picture, were not
  seen.  At the turn of the century, which coincided with the first operation of
  the PEPII/BaBar and KEKB/Belle ``$B$-factory'' experiments, it was generally thought
  that one of the tasks for early twenty-first century experiments would be the fleshing out
  some of the remaining unassigned charmonium (and bottomonium) levels.

One of the big surprises from the $B$-factory experiments was the discovery of mesons
with decay final states that include a $c$- and a $\bar{c}$-quark that cannot
be assigned to any of the remaining unassigned levels of the charmonium spectrum.
The first of these {\it charmoniumlike}\footnote{
  {\it Charmoniumlike} is used to designate states that
   appear to contain a $c$- and $\bar{c}$-quark but have properties that do not match
   expectations for a $\ccbar$ charmonium level.}
states to be observed was the  $X(3872)$
that was seen by Belle as a distinct narrow peak in the $\pipi\jp$ invariant mass
distribution in $B\rt K\pipi\jp$ decays~\cite{Choi:2003ue} (see Fig.~\ref{fig:x3872}).
The $\jp$ in its decay final state is a clear indication that the $X(3872)$, whatever
it is, must contain a $c$- and $\bar{c}$-quark.\footnote{
  The large mass of the $c$-quark ensures
  that the probability for the production of a $\ccbar$ pair from the vacuum during
  the quark to hadron fragmentation process is negligibly small.}
Although the original Belle report was
based on only $\sim$36~signal events, it established two properties of the $X(3872)$
that ruled against its interpretation as a two-quark, $\ccbar$ charmonium
state:\\
{\bf  i)}~its mass, reported at that time to be $M_{X(3872)}=3872.0 \pm 0.8$~MeV and
shown as a green horizontal line in Fig.~\ref{fig:charmonium}, was a poor match to
expectations for any of the unassigned $\ccbar$ charmonium states at that time;\\
{\bf ii)}~the $\pipi$ invariant mass peaked near the
$M_{X(3872)}-m_{\jp}\approx 775$~MeV kinematic boundary, consistent with an
$X(3872)\rt\rho\jp$, $\rho\rt\pipi$ decay chain. All charmonium states are
isoscalars and the $\rho$-meson is an isovector; if the $X(3872)$ is a charmonium
state, its decay to $\rho\jp$ would be a suppressed isospin-violating process
and an unlikely  discovery mode.

A third striking feature of the $X(3872)$ that was also noted in ref.~\cite{Choi:2003ue}
is that its mass is indistinguishable from the $\dz\dstrzbar$ mass threshold, which,
in 2003, was known to be $m_{\dz} + m_{\dstrz}=3871.1\pm 1.1$~MeV~\cite{Hagiwara:2002fs}
[$M_{X(3872)}-(m_{\dz}+m_{\dstrz}) =0.9 \pm 1.4$~MeV].\footnote{
  With 2018 PDG values ($M_{X(3872)}=3871.69 \pm 0.17$~MeV and
  $m_{\dz} + m_{\dstrz}=3871.70\pm 0.10$~MeV~\cite{Tanabashi:2018oca}),
   $M_{X(3872)}$ and the $\dz\dstrzbar$ mass threshold are even closer:
  [$M_{X(3872)}-(m_{\dz}+m_{\dstrz}) =-0.01 \pm 0.20$~MeV].}
This suggested that there is a close relationship between the $X(3872)$ and the $\dz\dstrzbar$
meson system.  In fact, two weeks after
Belle posted its first (preliminary) $X(3872)$ results in August 2003~\cite{Abe:2003hq},
T\"{o}rnqvist posted a note~\cite{Tornqvist:2003na} that identified it as a composite
deuteronlike $D\dstrbar$ state that he had predicted ten years earlier and
called a ``deuson''~\cite{tornqvist:1993ng}.  He predicted: its quantum numbers to be
$J^{PC}=0^{-+}$ or $1^{++}$; a width of order 50~keV; and a strong decay mode to be
$\dz\dzbar\piz$ via $\dz\dstrzbar$. What we now know about the $X(3872)$ aligns well
with T\"{o}rnqvist's predictions: LHCb established its $J^{PC}$ to be unambiguously
$1^{++}$~\cite{Aaij:2013zoa,Aaij:2015eva}; Belle placed an upper limit on its width of
1.2~MeV~\cite{Choi:2011fc}; and both Belle \& BaBar have reported that
$X(3872)\rt \dz\dzbar\piz$ is the dominant decay
mode~\cite{Gokhroo:2006bt,Aubert:2007rva,Adachi:2008sua}, with a branching fraction
that is greater than 40\%~\cite{Tanabashi:2018oca}.

The proximity of $M_{X(3872)}$ to the $\dz\dstrzbar$ mass threshold and the
plausibility of T\"{o}rnqvist's arguments encouraged us to believe that the $X(3872)$
was the harbinger of a new spectroscopy of open-charmed meson-meson molecules bound
by nuclear-physics-like forces, as first advocated in
1976~\cite{Voloshin:1976ap,Bander:1975fb,DeRujula:1976zlg}. So, in addition to
filling some of the gray boxes in Fig.~\ref{fig:charmonium} with bona-fide
$\ccbar$ states, my colleagues and I expected to spend the first few decades of
the twenty-first century establishing a new spectroscopy
of deuteron-like $D^{(*)}\bar{D}^{(*)}$ molecular states.

\section{What are they? ... or, better, what aren't they?}
\label{reality}
\begin{figure}[htb]
\begin{minipage}[t]{65mm}  \includegraphics[height=0.85\textwidth,width=0.85\textwidth]{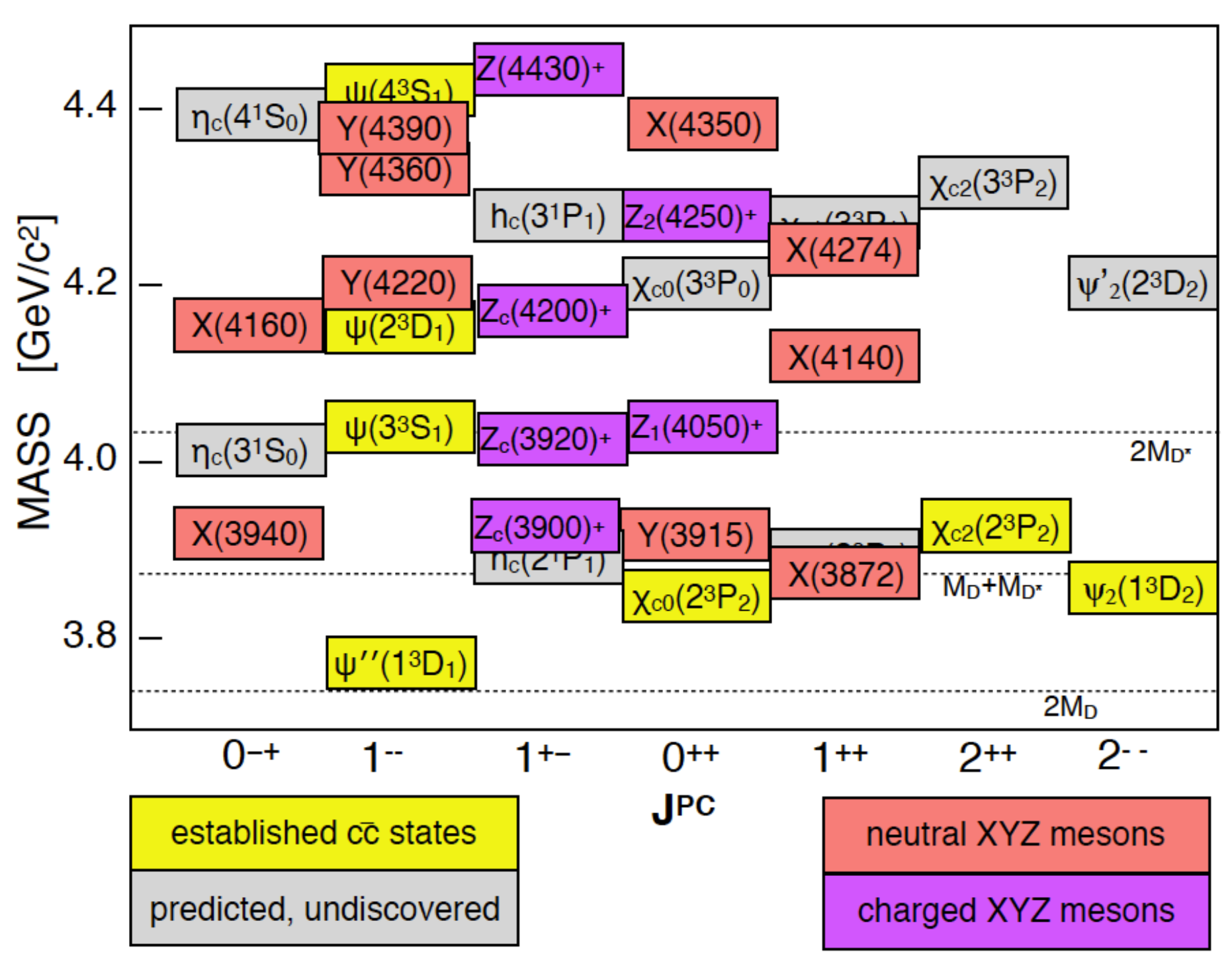}
\caption{\footnotesize The above open-charm-threshold charmonium \& charmoniumlike spectrum in 2018.
}
\label{fig:charmonium+xyz}
\end{minipage}
\begin{minipage}[t]{22mm}
\end{minipage}
\begin{minipage}[t]{70mm}
  \includegraphics[height=0.75\textwidth,width=0.85\textwidth]{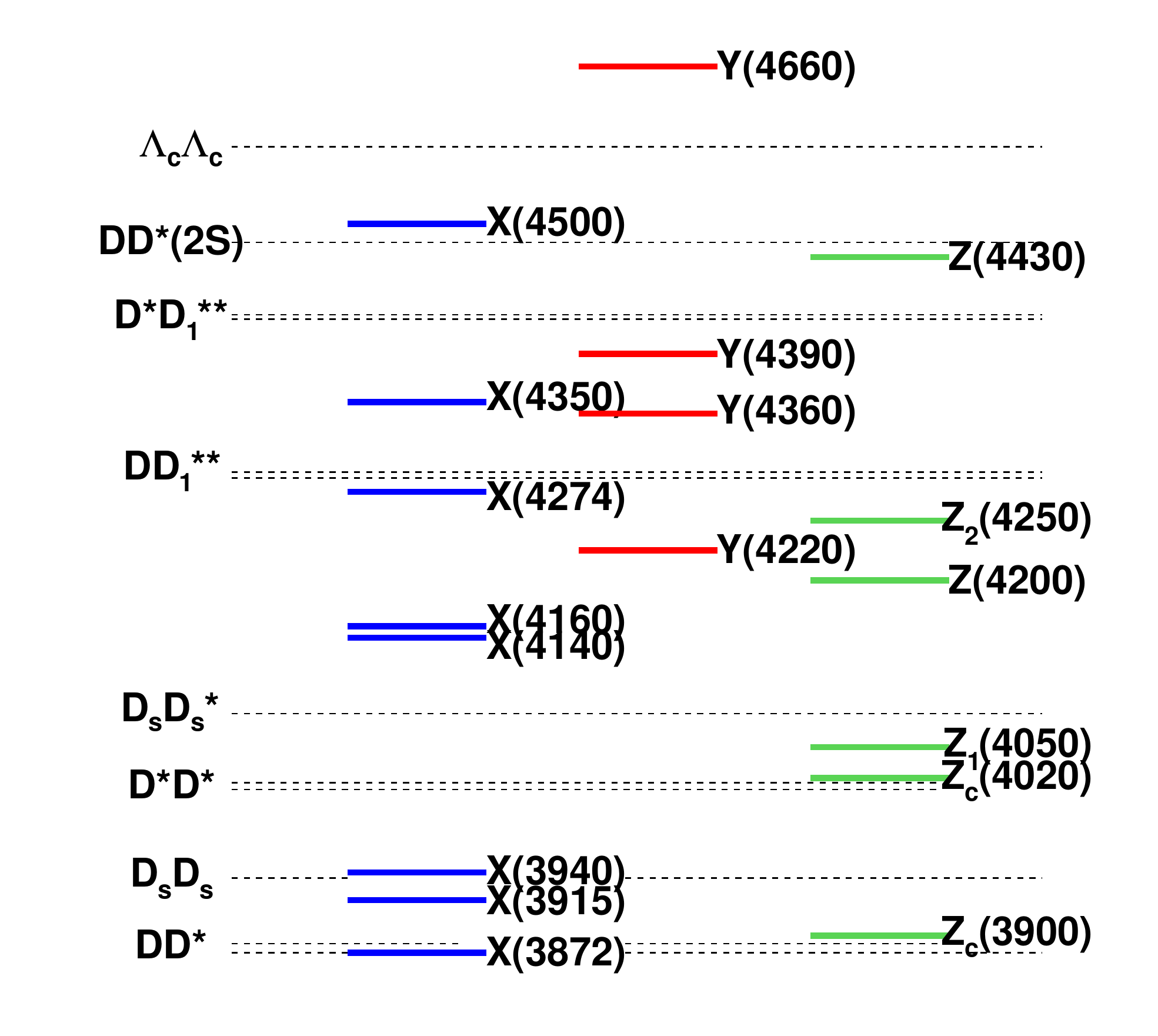}
%\end{minipage}\hspace{\fill}
  \caption{\footnotesize $XYZ$ meson masses compared with charmed meson pair thresholds.
  }
\label{fig:thresholds}
\end{minipage}\hspace{\fill}
\end{figure}

Sure enough, as the $B$-factory programs unfolded, and BESIII started up,
additional $\ccbar$ charmonium states were found,\footnote{
  The $\chiczerop$, $\chictwop$ and $\psi_2(1^{3}D_2)$.}
along with a larger number of charmoniumlike states, both neutral and charged,
as indicated in Fig.~\ref{fig:charmonium+xyz}.  The properties of these states, which
are collectively known as the $XYZ$ mesons,
have been extensively reviewed (see, for example, ref.~\cite{Olsen:2017bmm}) and
are generally well known.  What is not well known is {\it what they are}, and this has
turned out to be a very challenging issue. Here I address
a more modest question: {\it what aren't they?}

Proposed theoretical models for these new states include:\\
{\bf molecules:}~~loosely bound deuteron-like meson-meson structures;\\
{\bf QCD tetraquarks:}~~colored quark ($[cq_i]$) and diantiquark ($[\bar{c}\bar{q}_j]$)
  configurations ($q_i=u,d,s$) tightly bound by the exchange of colored gluons;\\
{\bf charmonium hybrids:} a $\ccbar$ pair plus an excited ``valence'' gluon (and electrically neutral);\\
{\bf threshold effects:} enhancements caused by threshold cusps,
  rescattering processes, etc.;\\
{\bf hadrocharmonium:} a colorless hadron cloud of light quarks \& gluons,
  bound to a $\ccbar$ charmonium core state via van-der-Waals forces.\\

Here I briefly discuss each of these possibilities, with emphasis on their experimental
consequences. I restrict the discussion to six candidate $XYZ$ mesons that are experimentally
well established and whose $J^{PC}$ values are known: {\it i.e.,} the isospin zero $X(3872)$,
$X(3915)$~\cite{Abe:2004zs,Uehara:2009tx} and $Y(4220)$~\cite{Aubert:2005rm},\footnote{
  Commonly known as $Y(4260)$, but whose mass has recently been measured to be
 $4222 \pm 3$~MeV by BESIII~\cite{Ablikim:2016qzw}.}% (See Fig.~\ref{fig:pipijp} below.)}
and the isospin one $Z_c(3900)$~\cite{Ablikim:2013mio,Liu:2013dau}, $Z_c(4020)$~\cite{Ablikim:2013wzq}
and $Z(4430)$~\cite{Choi:2007wga}.

\subsection{Molecules:}
The expected properties of a deuteronlike molecular state are conveniently listed by Karliner
and Skwarnicki in the context of remarks about Pentaquarks in the PDG 2018
report~\cite{Tanabashi:2018oca}:\\
\noindent
{\bf a)} mass near the constituent meson-meson threshold
     and $J^{PC}$ consistent with an $S$-wave;\\
{\bf b)} narrow despite the large phase-space for $\ccbar$~+~pion(s) decays;\\
{\bf c)} branching fraction for meson-meson ``fall-apart'' decay larger
     than that for  $\ccbar$~+~pion(s);\\
{\bf d)} not a pseudoscalar-pseudoscalar, for which single-pion exchange
     is not allowed;\\
{\bf e)} wider than either of its constituents.\\
I take ``near the constituent meson-meson threshold,'' to mean
$BE\lesssim m_{\pi}^2/2\mu \approx 10$~MeV (for reduced mass $\mu\approx m_D/2$), corresponding
to an rms meson-meson separation $d_{\rm rms}\gtrsim m_{\pi}^{-1}$.

Figure~\ref{fig:thresholds} shows how the measured $XYZ$ meson masses compare with the
%various $D^{(*)}\bar{D}^{(*)(*)}$
charmed-particle/anticharmed particle mass thresholds below 4600~MeV. No clear pattern of
$XYZ$ states favoring thresholds is evident.  The $Z_c(3900)$ and $Z_c(4020)$ are above, but within
$\sim$10~MeV, of the $D\bar{D}^*$ and $D^*\bar{D}^*$ thresholds, respectively and qualify
as unbound, virtual meson-meson states.  The $X(3915)$ is $\approx$100~MeV below $2m_{D^*}$
and 18~MeV below $2m_{D_s}$. The binding energy required for $D^*\bar{D}^*$ molecule
is too high;\footnote{In ref.~\cite{Molina:2009ct}, the $X(3915)$ is interpreted
  as a (mostly) $D^*\bar{D}^*$ system tightly bound by vector-meson exchange and a vector-vector
  contact term. However, there is no independent evidence for the existence of the proposed
  binding mechanism, and the predicted accompanying $1^{+-}$ and $2^{++}$ mesons
  have not been seen.}
$D_s\bar{D_s}$ is a disqualified pseudoscalar-pseudoscalar combination
(this is discussed in ref.~\cite{Li:2015iga}). Some authors interpret the $Y(4220)$
as an $S$-wave $D\bar{D}_1(2420)$ molecule, but provide no explanation for the
$\approx$65~MeV binding energy this would imply~\cite{Cleven:2013mka}. The mass of the $Z(4430)$,
now established to be $4478\pm 18$~MeV, is equal within errors to
$m_{D}+M_{D(2600)}\simeq 4480$~MeV, where the $D(2600)$ is a candidate for the $D^*(2S)$ radial
excitation of the $D^*$ that was reported by BaBar~\cite{delAmoSanchez:2010vq}. However,
$\Gamma_{Z(4430)}=181 \pm 31$~MeV, and $\Gamma_{D(2600)}=93 \pm 15$~MeV, and one wonders
if the concept of molecule applies to objects with such short lifetimes.

\subsubsection{The X(3872) as a molecule?}
Although the $X(3872)$ is often considered to be the prototypical meson-meson
molecule, this may not be the case.  Its decay to $D^0\bar{D}^{*0}$ means that its $S$-wave
$D\bar{D}^*$ coupling, $g_{D\bar{D}^*}$, is non-zero and, since its mass is very near
$m_{\dz}+m_{\dstrzbar}$, the effects of $g_{D\bar{D}^*}$ get strongly amplified by the nearly
divergent $[M_X-(m_{\dz}+m_{\dstrzbar})+k^2/2\mu ]^{-1}$ propagator that occurs in coupled-channel
calculations. So, whatever its underlying nature may be, the near equality of the $X(3872)$
mass with $m_{\dz}+m_{\dstrzbar}$
\footnote{It is not known if
  this near equality is the result of some dynamics or just a coincidence.}   
will make it behave like a $D\bar{D}^*$ molecule~\cite{Braaten:2007dw,Coito:2012vf}. Detailed
calculations show that coupled-channel effects are more important than
meson-meson binding~\cite{Takeuchi:2014rsa}. 

\subsection{QCD tetraquarks}
Since the diquark and diantiquark in a QCD tetraquark are bound by the color confining
force, the binding energies are technically infinite and strong mass affinities
for meson-meson thresholds are not expected; just about any mass and many $J^{PC}$ values
can be accommodated. Thus, in the absence of a specific model, any charmoniumlike meson
state with quantum numbers consistent with a $[cq_i][\bar{c}\bar{q}_j]$ arrangement
can be explained as a QCD tetraquark.  On the other hand, since the QCD color force is flavor
blind and the same for $[cu]$, $[cd]$ and $[cs]$ diquarks (and diantiquarks), QCD tetraquarks
should form $SU(3)$
nonets~\cite{Luo:2017eub}. However, other than the $Z_c(3900)$ and $Z_c(4020)$ isospin
partners, none of the expected nonet partner states have been seen.  This may reflect a lack
of experimental sensitivity, but, in cases where experimentally verifiable
predictions have been made~\cite{Luo:2017eub,Maiani:2004vq}, the expected partner
particles have not been found~\cite{Aubert:2004zr,Choi:2011fc}.

The $X(3915)$, which is an unlikely candidate for a molecule (see above)
and too light to be a charmonium hybrid (see below), is, by default, a
candidate for an $[cs][\bar{c}\bar{s}]$ QCD tetraquark
state~\cite{Lebed:2016yvr}. In this case, its quark content would be better matched to
$\eta\eta_c$ than to $\omega\jp$,\footnote{The $\eta$'s $|s\bar{s}>$ and
  $|u\bar{u}+d\bar{d}>/\sqrt{2}$ contents are nearly equal~\cite{Bramon:1997va};
  the $\omega$'s $|s\bar{s}>$ content is nearly zero~\cite{Epele:2002qe}.}
and one would na{\" i}vely expect the partial decay width for $X(3915)\rt\eta\eta_c$
to be substantially larger than that for the $X(3915)\rt\omega\jp$ discovery channel.
A Belle search for $X(3915)\rt\eta\eta_c$ saw no signal and set a upper
limit $\Gamma_{X\rt\eta\eta_c} < 1.5\times \Gamma_{X\rt\omega\jp}$~\cite{Vinokurova:2015txd},
which is not encouraging for a QCD tetraquark assignment.

\subsection{Charmonium hybrids}

Of the six $XYZ$ mesons that we are considering, only the $X(3872)$, $X(3915)$ and $Y(4260)$
are electrically neutral and viable candidates for $\ccbar$-gluon charmonium hadrons. The
strongest positive indication of a charmonium hybrid would be exotic spin-parity
quantum numbers, {\it e.g.,} $J^{PC}$ values that cannot be accessed by a
fermion-antifermion pair, but could be formed by a $\ccbar$-gluon system.
Examples would be $J^{PC}=0^{--}$, $0^{+-}$, $1^{-+}$ or $2^{+-}$ mesons. However,
all $XYZ$ meson candidates reported to date have non-exotic $J^{PC}$ values.
Another charmonium hybrid characteristic would be a preference to decay to a
$D^{(*)}\bar{D}^{**}$ pair, {\it i.e.,} an $S$-wave $c\bar{q}_i$ meson plus
$P$-wave $\bar{c}q_j$ antimeson ($q_i=u,d$), or vice-versa~\cite{Isgur:1985vy}.
However the only distinctively narrow  and relevant $P$-wave $c\bar{q}$ meson is
the $D_1(2420)$, and the $D\bar{D}_1(2420)$ decay channel is energetically inaccessible
to all three states. 

The Hadron Spectrum Collaboration (HSC) reported charmonium and charmonium-hybrid
mass values calculations performed on two lattice volumes with a pion mass
$\approx$400~MeV~\cite{Liu:2012ze}. Their lightest $1^{++}$ hybrid mass value is
$\approx$4400~MeV, more than 500 MeV too high for an $X(3872)$ assignment, and
their lightest $0^{++}$ hybrid mass is $\approx$4480~MeV, an equally poor match to
the $X(3915)$. On the other hand, their mass value for the lightest $1^{- -}$ hybrid is
$\approx$4380~MeV, and consistent with the $Y(4220)$ mass within the $\sim$100~MeV
precision that characterizes their calculation.\footnote{The HSC-calculated masses for the
  $\chi_{c0}^{\prime}$ and $\chi_{c2}^{\prime}$ charmonium states
  are also high by about $100$~MeV.}
Thus, although there is no other strong evidence to back a charmonium-hybrid assignment
for the $Y(4220)$, there is nothing that rules it out.

\subsection{Threshold effects}

In coupled channel systems that involve an $S$-wave meson-meson system (the ``elastic
channel''), cusp-like peaks can be produced in other channels by purely kinematic
effects~\cite{Bugg:2011jr,Blitz:2015nra,Swanson:2015bsa} or by rescattering processes
with internal triangular loops~\cite{Chen:2013coa,Pakhlov:2014qva} that become singular
when the internal particles go on the mass shell~\cite{Landau:1959fi}. These peaks
occur at masses just above the relevant threshold and have narrow, but non-zero widths.
The $Z_c(3900)$, seen as $S$-wave $\pi\jpsi$ and $D\bar{D}^*$ mass peaks just above
the $D\bar{D}^*$ threshold, is a candidate for this kind of effect, as is the
$Z_c(4020)$, which is seen as $\pi h_c$ and $D^*\bar{D}^*$ mass peaks just above the
$D^*\bar{D}^*$ threshold.\footnote{The
  $Z_b(10,610)$ and $Z_b(10,650)$ ``bottomoniumlike'' mesons are
  seen as $\pi\Upsilon(nS)$ ($n=1,2,3$), $\pi h_b(mP)$ ($m=1,2$) and $B^{(*)}\bar{B}^*$
  mass peaks just above the $B\bar{B}^*$ and $B^*\bar{B}^*$ thresholds.}
An analysis of the $Z_c(3900)$~\cite{Guo:2014iya}
concluded that while a kinematic cusp just above the $D\bar{D}^*$ threshold can be
produced in the $\pi\jp$ mass distribution, this effect cannot produce a similarly narrow
peak in the elastic $D\bar{D}^*$ channel. Thus, according to ref.~\cite{Guo:2014iya},
BESIII's narrow $Z_c(3900)\rt D\bar{D}^*$ signal~\cite{Ablikim:2013xfr} establishes the
presence of a genuine meson-like pole in the $D\bar{D}^*$ $S$-matrix. Similar considerations
obtain for the $Z_c(4020)$ and its $D^*\bar{D}^*$ decay mode~\cite{Ablikim:2013emm}.
A more general discussion of the theoretical issues is provided in ref.~\cite{Szczepaniak:2015hya}.

\subsection{Hadrocharmonium}

For conventional charmonium states that are above the open-charmed meson pair threshold,
branching fractions for ``fall-apart'' decays to charmed meson pairs are 2~or~3~orders
of magnitude higher than decays to hidden charm states.  On the other hand, most of the
$XYZ$ mesons were discovered via their hidden charm decay modes, which, in contrast to
ordinary charmonium states, have branching fractions that are within one order of
magnitude of those for fall-apart
modes.  The hadrocharmonium mode was proposed to account for this. In this model, a
compact color-singlet $\ccbar$ charmonium core state is embedded in a spatially
extended “blob” of light hadronic matter. These two components interact via a QCD
version of the van der Waals force~\cite{Dubynskiy:2008mq}. In the case of the
$Y(4220)$,  this core state was taken to be the $\jp$. Since the $\jp$ is present
in its constituents, the $Y(4220)$ naturally prefers to decay to final states that
include it, such as the $Y(4220)\rt\pipi\jp$ discovery mode.

\begin{figure}[htb]
\begin{minipage}[t]{65mm}  \includegraphics[height=0.75\textwidth,width=0.95\textwidth]{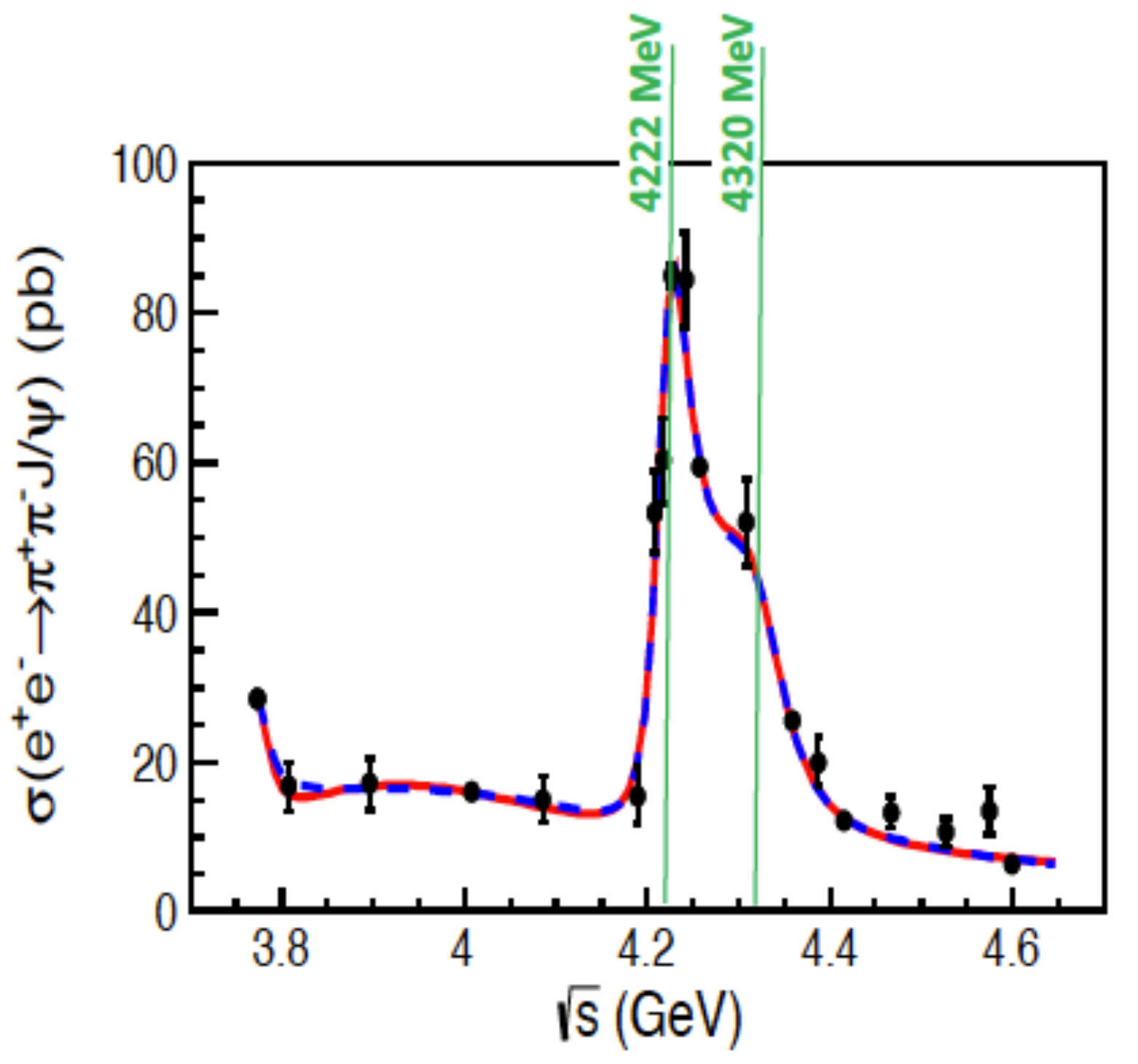}
  \caption{\footnotesize BESIII $\sigma(\ee\rt\pipi\jp)$ measurements~\cite{Ablikim:2016qzw}
    that show that the $Y(4260)$ is actually two peaks with masses $4222\pm 3$~MeV
    and $4320\pm 13$~MeV. The $Y(4220)\rt\pipi\jp$ peak cross section is $85\pm 6$~pb.
}
\label{fig:pipijp}
\end{minipage}
\begin{minipage}[t]{22mm}
\end{minipage}
\begin{minipage}[t]{70mm}
  \includegraphics[height=0.65\textwidth,width=0.85\textwidth]{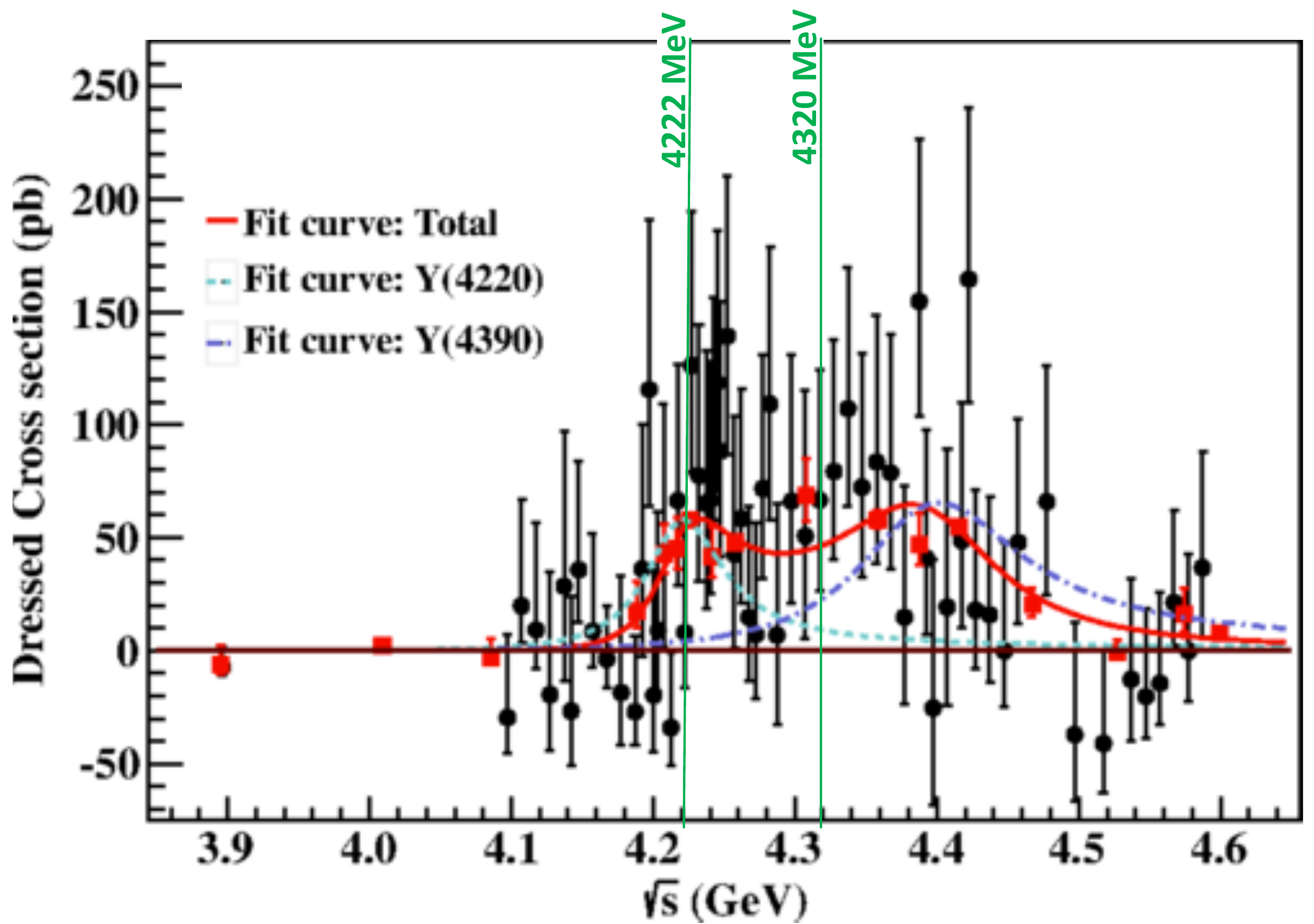}
%\end{minipage}\hspace{\fill}
  \caption{\footnotesize BESIII $\sigma(\ee\rt\pipi h_c)$  measurements~\cite{BESIII:2016adj}.
    Here there is a strong $Y(4220)$ signal but the $Y(4320)$ is absent.
     The $Y(4220)\rt\pipi h_c$ peak cross section is $55\pm 9$~pb.
  }
\label{fig:pipihc}
\end{minipage}\hspace{\fill}
\end{figure} 

Precision BESIII measurements of  $\sigma(\ee\rt\pipi\jp)$, shown in Fig.~\ref{fig:pipijp},
revealed two peaks in the $\sqrt{s}=4260$~MeV region: the $Y(4220)$ and
$Y(4320)$~\cite{Ablikim:2016qzw}. 
Measurements of $\sigma(\ee\rt\pipi h_c)$ (Fig.~\ref{fig:pipijp})~\cite{BESIII:2016adj},
show that the $Y(4220)\rt \pipi h_c$ decay
%\footnote{The $h_c$ is the lowest singlet $P$-wave ($1^1P_1$) $\ccbar$ state.}
branching fraction is comparable to that for $\pipi\jp$. Since the $\ccbar$
is in a spin-singlet state in the $h_c$ and a spin triplet state in the $\jp$, the $\ccbar$
core in the hadrocharmonium version of the $Y(4220)$ should be one or the other, but not a
mixture of the two. However, the $Y(4220)$ itself could be a mixture of two
hadrocharmonium states, one with an $h_c$ core and the other with a $\jpsi$
core~\cite{Li:2013ssa}. This implies the existence of two $Y(4220)$-like states with
orthogonal $h_c$-$\jp$ mixtures.  The obvious candidate for the second state is the
$Y(4360)$~\cite{Ablikim:2016qzw}, but there is no sign of it in the $\sigma(\ee\rt\pipi h_c)$
measurements shown in Fig.~\ref{fig:pipihc},\footnote{The second peak in Fig.~\ref{fig:pipihc} is
  at $4392 \pm 7$~MeV and quite distinct from the $4320$~MeV structure in Fig.~\ref{fig:pipijp}.} 
as would be expected for a $\jp$-$h_c$ mixture orthogonal to the $Y(4220)$.  Even though
hadrocharmonium was originally proposed as an explanation for the properties of the $Y(4220)$,
it has trouble explaining BESIII's $Y(4220)rt\pipi\jp$ and $\pipi h_c$ measurements.
Recently, BESIII reported observation of $X(3872)\rt\piz\chi_{c1}$ with a
(preliminary) branching fraction that is
$(0.9\pm 0.3)\times {\mathcal B}(X(3872)\rt\pipi\jp)$~\cite{Ryan:2018ab}.
This implies a similar dilemma for a hadrocharmonium interpretation for the $X(3872)$.

\section{No single size fits all}

Table~\ref{tab:compare} summarizes the above discussion, where the red entries indicate
assignments that are ruled out and the blue ones designate the best of the
remaining possibilities for each meson under consideration. Possiblities that the
$Z_c$ states may be threshold effects are indicted in olive (and not red) because,
in spite of the arguments in ref.~\cite{Guo:2014iya}, the match between the properties
of these states (and the similar $Z_b$ states) and expectations for kinematically
induced peaks ({\it i.e.}, masses just above threshold, similar widths, not seen
in $B$-meson decays, etc.) is so uncanny, I think more information is needed before
they can be conclusively ruled out. The black question marks reflect my lack of knowledge.

While red entries indicate assignments that I consider ruled out for reasons given
above, blue entries are blue mainly by default.  Other than that for the $X(3872)$, blue
assignments are not strongly supported by experimental evidence, but are not ruled out
either. Establishing  what the $XYZ$ mesons {\it are} will require more experimental
and theoretical investigation.

\begin{table}[htb]
\centering
\footnotesize{
  \caption{Comparison of meson properties with model expectations. The text describes the color code.}
\label{tab:compare}
% For LaTeX tables you can use
\begin{tabular}{lccccc}
\hline\hline
state     &   molecule?     &  tetraquark? & charmonium & kinematic & hadro- \\
          &                 &              &  hybrid?   &  effect?  &charmonium?\\
\hline %----------------------------------------------------------------------------------------------------
\hline %----------------------------------------------------------------------------------------------------
$X(3872)$ &  {\color{blue} coupled-channel} & {\color{red} partner} &  {\color{red} m$\approx$500 MeV}
&  {\color{red} width too} &  {\color{red} decays to}     \\
          &   {\color{blue} system;}        & {\color{red} states not}      &    {\color{red} too low}
& {\color{red} narrow}     & {\color{red} $\gamma\jp$ \& $\gamma\chi_{c1}$}  \\
          &   {\color{blue} not a deuson}   & {\color{red} found}   &
&               &                                        \\                                   
\hline %----------------------------------------------------------------------------------------------------
$X(3915)$ & {\color{red} $\pi $-exchange}    & {\color{blue}$\eta\eta_c$ decay} &  {\color{red} m$\approx$500 MeV}
&  {\color{red} no nearby}  &???   \\
          & {\color{red} forbidden}          &   {\color{blue} not seen}        & {\color{red} too low}
&  {\color{red}threshold}  &       \\
\hline %----------------------------------------------------------------------------------------------------
$Y(4220)$ &    {\color{red} $D\bar{D}_1(2420)$} &   ???         &   {\color{blue} possible}
& {\color{red} no nearby}  &{\color{red} decays to}  \\
          & {\color{red} BE$\approx$65 MeV}     &               &
&  {\color{red} threshold}         &   {\color{red}  $\pipi\jp$}    \\
          &  {\color{red} -too high-}             &               &           
&                                  &   {\color{red} \& $\pipi h_{c}$}  \\
\hline %----------------------------------------------------------------------------------------------------
$Z_C(3900)$& {\color{blue} $D\bar{D}^*$}   &  ???        & {\color{red} Isospin=1}
& {\color{olive} possible?}  &   ???           \\  
          &  {\color{blue} virtual state?}                 &         &
&                            &                 \\
\hline %----------------------------------------------------------------------------------------------------
$Z_c(4020)$& {\color{blue} $D^*\bar{D}^*$} &  ???       & {\color{red} Isospin=1}
&   {\color{olive} possible?} &     ???         \\
           &  {\color{blue} virtual state?}          &          &
&                             &                 \\
\hline %----------------------------------------------------------------------------------------------------
$Z(4430)$ &  {\color{blue} too wide for} & ???        & {\color{red} Isospin=1}
&  {\color{red} too wide}     &     ???        \\
          &  {\color{blue} a  $D\bar{D}^*(2P)$}       &             &
&                              &               \\
&  {\color{blue} molecule?}     &            &
&                              &               \\
\hline %----------------------------------------------------------------------------------------------------
\hline %--------------- -------------------------------------------------------------------------------------

\end{tabular}

}
\end{table}

I conclude that no single one of the models addressed above can satisfactorily
explain all the results.  If we are ever to have a coherent, comprehensive understanding
of the $XYZ$ particles, a new idea is needed. Otherwise we will be left with
an (unsatisfactory) menu of different models with column A for some states, column B for
others, etc.

\section{Acknowledgements}
I congratulate Charm2018 organizers for arranging an interesting and provocative meeting,
and thank them for providing me the opportunity to present these remarks.  The work is
supported by the Chinese Academy of Science President's International Fellowship Initiative.
I thank Chengping Shen for helpful comments.

\bibliography{charm2018_olsen}

% BibTeX or Biber users please use (the style is already called in the class, ensure that the "woc.bst" style is in your local directory)

%
% Non-BibTeX users please use
%
%\begin{thebibliography}{}
%
% and use \bibitem to create references.
%
%\bibitem{RefJ}
% Format for Journal Reference
%Journal Author, Journal \textbf{Volume}, page numbers (year)
% Format for books
%\bibitem{RefB}
%Book Author, \textit{Book title} (Publisher, place, year) page numbers
% etc
%\end{thebibliography}

\end{document}